\documentclass[a4paper, 10pt, twocolumn, superscriptaddress]{revtex4-2}
\usepackage{amsmath}
\usepackage{amsfonts}
\usepackage{graphicx}
\usepackage{setspace}
\usepackage{xcolor}
\usepackage{comment}
\usepackage{ulem}

\usepackage{soul}

\def\EE{{\cal E}}

\def\VV{{\cal V}}

\begin{document}

\title{Correlation distances in social networks}
\author{P\'adraig MacCarron}
	\affiliation{MACSI, Department of Mathematics and Statistics, University of Limerick, Limerick, Ireland}
	\affiliation{Centre for Social Issues Research, University of Limerick, Limerick, Ireland}
\author{Shane Mannion}
	\affiliation{MACSI, Department of Mathematics and Statistics, University of Limerick, Limerick, Ireland}
\author{Thierry Platini}
\affiliation{Applied Mathematics Research Centre, Coventry University, Coventry, England}

\begin{abstract}
In this work we explore degree assortativity in complex networks, and extend its usual definition beyond that of nearest neighbours. We apply this definition to model networks, and describe a rewiring algorithm that induces assortativity. We compare these results to real networks. Social networks in particular tend to be assortatively mixed by degree in contrast to many other types of complex networks. However, we show here that these positive correlations diminish after one step and in most of the empirical networks analysed. Properties besides degree support this, such as the number of papers in scientific coauthorship networks, with no correlations beyond nearest neighbours. Beyond next-nearest neighbours we also observe a diasassortative tendency for nodes which are three steps away from each other, indicating that nodes at that distance are more likely to be different than similar. 
\end{abstract}
\maketitle

\section{Introduction}

In recent decades, many quantities have been introduced to shed light on the structural properties of complex networks. Of particular importance is degree assortativity, which provides a measure of the correlations in the degree of neighbouring nodes. Its importance was underlined due to it distinguishing the structure of social networks, and other types of complex networks~\cite{NewmanPark} (though this is found to not always be the case for online social networks~\cite{HuWang}). Studies of assortativity have focused on nearest neighbour correlations, since neighbours interact directly only with their nearest neighbours, by construction. 

By drawing parallels with spin systems, such as that described by the Ising model, one can argue that system components may exert influence indirectly, beyond their immediate neighbourhood. This is captured in the idea of \textit{correlation length}. In the Ising model, while an individual spin has no direct interaction with more distant spins, a re-orientation of a spin can influence its neighbours causing a disturbance to propagate over a large area of the lattice, modelling the observed behaviour of ferromagnets. Regions separated by more than this characteristic distance can be thought of as essentially independent~\cite{Wilson}. Here we examine whether degree correlations in complex networks exist beyond those found between nearest neighbours.

In this work we study correlations on undirected, unweighted networks as a function of network distance. The most elementary feature of a network node is its \textit{degree}, defined as the number of edges adjacent to that node. The network-wide correlation between degrees of neighbouring nodes is commonly known as \textit{assortativity}~\cite{Newman2002}. A network with positive degree correlations between nodes is called assortative, and with negative correlations, disassortative. We use the shortest path length between two nodes to represent the distance, when this is greater than one we are moving beyond the traditional nearest neighbours.

In social networks, where node degree represents the number of friends that an individual has, assortativity indicates that popular individuals tend to have popular friends, and unpopular individuals unpopular friends. This notion is related to that of \textit{homophily}, whereby individuals associate with people similar to themselves. This leads to a tendency for individuals to associate mostly with others of a similar race or ethnicity, age, religion or interest~\cite{McPherson2001}.

Previous studies have suggested that nodes' influence may extend beyond the immediate neighbourhood in social networks (e.g.~\cite{C+F2007}). It has even been claimed that individuals can have up to ``three degrees of influence'' on other nodes in a network~\cite{C+F2009}; though serious questions have been raised about the methods of these studies~\cite{Lyons2011}. Here we use the assortativity at different distances to test this. In particular, we use a co-authorship network sampled at three distinct time points. If nodes are expected to influence each other at a distance, na\"ively we might expect the beyond nearest neighbour correlations (between both degrees and number of publications) to increase over time.

Related works have examined how correlations vary with distance, typically using the length of the shortest path between two nodes as a measure of the distance between them. Arcagni et al.~\cite{Arcagni} apply three different methods to understand higher-order assortativity; a random-walk based approach, a path-based approach and a degree-based path approach.

Another theoretical approach is taken in by Fujiki et al.~\cite{Fujiki}. Here the authors look at several joint and conditional probability distributions relating node degrees and the shortest path lengths that separate them. These methods are more focused on the mean-field scenario rather than correlations in empirical datasets as we are here. Mizutaka et al.~\cite{Mizutaka} observe a critical value for the average degree of an Erd\H{o}s-R\'enyi random graph, above which negative long distance correlations emerge. Finally, Mayo et al.~\cite{Mayo} study three social networks using a slightly different approach. They look at the average degrees of nodes at the end of paths of a certain length from source nodes and plot this versus the source node degree to view correlations. However, they only study correlation between nodes separated by paths of length four or less on three social networks, concluding that more research is necessary. 

In this paper, we study path lengths up to the diameter of our graphs for 16 empirical social networks. In the following section we outline the theoretical framework for our correlation measure, and introduce an expression to quantify assortativity as a function of distance. We then apply this to 16 social networks to observe correlation behaviour beyond nearest neighbours. We then perform simulations on random graphs, configuration model variants of each empirical network, as well as generated assortative networks, to attempt to generalise the results we observe.
\section{Theory}
In this section we define a number of quantities centered around the concept of assortativity. We start by defining a graph or network as an ordered pair $G = (\VV, \EE)$, where $\VV$ is the set of nodes and $\EE$ is the set of edges. The number of nodes and edges in a network are written $N=|\VV|$ and $L=|\EE|$, respectively. Denoting the adjacency matrix by $A$, with elements $A_{ij}$, the degree $k$ of node $i$ is given by $k_i = \sum_jA_{ij}$. The mean degree of the network is given by $\langle k \rangle = \sum_i k_i/N $. Although our results may be extended to directed graphs, for the purposes of this study, we focus solely on undirected networks, meaning that $A_{ij} = A_{ji}$. It follows that $2L=\sum_{i,j}A_{ij}=\sum_{i}k_{i} = \sum_k k p_k$, where $p_k$ is the probability that a randomly chosen node has degree $k$, and so the final term is just $N$ times the mean degree of the network.

The shortest path between nodes $i$ and $j$ is the path from $i$ to $j$ that traverses the minimal number of edges. We define the length of this path by $\lambda_{ij}$, noting that the path itself is often not unique. It is common for networks to consist of disconnected components, or sets of nodes that are mutually reachable by traversing edges. When this is the case, there exists no path between nodes $i$ and $j$ belonging to distinct components and by convention we set $\lambda_{ij}=\infty$. To define the average path length $\ell$ it is convenient to introduce $\Gamma$, the  number of components of a graph and to define $C_m$, with $m\in\{1,2,\hdots,\Gamma\})$, as the $m^{\rm th}$ component, which contains $n_m$ nodes. Summing over all pairs of a connected component, the average path length is given by
\begin{equation}
\ell = \sum_{m=1}^\Gamma  \frac{1}{n_m(n_m-1)} \sum_{ij\in{\cal C}_m} \lambda_{ij}. 
\label{eqn:average_path_length}
\end{equation}
The greatest shortest path length $\ell_{\rm{max}}$ is known as the diameter of the network.
\subsection{Assortativity}
Consider a randomly chosen edge. We introduce $\kappa$ and $\nu$ to represent the degrees at either end of this edge. When we discuss the degree of a node $i$, we use the standard notation $k_i$ The joint probability that an edge has a node of degree $\kappa$ at one end and a node of degree $\nu$ at the other is given by
\begin{equation}
 \Phi(\kappa,\nu) = \dfrac{1}{2L} \sum_{i,j} A_{ij}\delta_{k_i \kappa} \delta_{k_j \nu},
 \label{eqn:Phi}
\end{equation}
where $\delta_{k_i \kappa}$ is the Kronecker delta which is 1 when $k_i=\kappa$ and 0 otherwise.
We denote the average degree of the node at the end of an edge by $\mathbb{E}[\kappa]$. We stress that an average over edges is not the same as an average over nodes and to keep this distinction clear we use the notation $\mathbb{E}[\kappa]$~to represent the former and the notation $\langle k \rangle$ to represent the latter. To see how these quantities differ, note that in calculating $\mathbb{E}[\kappa]$, a node of degree $k$ will appear $k$ times in the summation.

The above quantities are, however, closely related. Further, note that since we have undirected graphs, $ \mathbb{E}[\nu] = \mathbb{E}[\kappa]$. Summing over $\nu$, the marginal probability is given by
\begin{equation}
    \phi(\kappa) = \sum_\nu \Phi (\kappa, \nu).
\end{equation}
Using this expression, it is possible to show that $\mathbb{E}[\kappa]$ and the mean degree over the nodes of the network are related by 
\begin{equation}
 \mathbb{E}[\kappa] = \sum_\kappa \kappa \phi(\kappa) = \frac{\sum_k k^2 p_k}{\sum_k k p_k},
 \label{eqn:assort_mean_deg}
\end{equation}
which is the second moment of the degrees divided by the mean degree (both moments taken over the set of nodes). Note that any network besides those which are $k$-regular will have positive degree variance. This implies that, except in the degree-regular case, $\mathbb{E}[\kappa]$ is greater than the mean degree of the nodes of the network. From this, if we select a node at random its degree will be on average $\langle k \rangle$. If we then follow a random edge from this node, the degree of the node at the other end will be $\mathbb{E}[\kappa]$ on average. This implies that nodes are, on average, connected to nodes with degrees higher than their own. This observation is commonly known as the friendship paradox~\cite{friendshipparadox}. Our starting point towards understanding degree correlations and assortativity is the expected value of the product of the degrees of the nodes at the end of an edge chosen uniformly at random, $\mathbb{E}[\kappa \nu] $, defined by
\begin{equation}
    \mathbb{E} [\kappa \nu] = \sum_{\kappa, \nu} \kappa \nu \Phi(\kappa, \nu) .
    \label{eqn:Ekq}
\end{equation}  
For uncorrelated, undirected graphs the probability distribution $\Phi(\kappa, \nu)$ can be factorized as
$\Phi(\kappa, \nu) =  \phi(\kappa) \phi(\nu)$.
The latter relation implies that in this special case
\begin{equation}
    \mathbb{E} [\kappa \nu] 
    = \mathbb{E}[\kappa] \mathbb{E}[\nu] =\mathbb{E}[\kappa]^2 = \mathbb{E}[\nu]^2 .
\end{equation}
It follows that a network is said to have positive correlations when $\mathbb{E} [\kappa \nu] > \mathbb{E} [\kappa] \mathbb{E} [\nu]$, and negative correlations when $ \mathbb{E} [\kappa \nu] < \mathbb{E} [\kappa]\mathbb{E}[\nu]$. 
We can now introduce the degree assortativity $r$ as
\begin{equation}
    r = \frac{\mathbb{E}[\kappa \nu] - \mathbb{E}[\kappa]\mathbb{E} [\nu]}{\sigma_\kappa^2},
    \label{eqn:assort1}
\end{equation}
which is in the form of the Pearson correlation coefficient~\cite{Newman2002}, with $\sigma_\kappa^2$ equal to the variance of $\kappa$,
\begin{equation}
    \sigma_\kappa^2 = \mathbb{E}[\kappa^2] - \mathbb{E}[\kappa]^2
\end{equation}
Normalising by the variance ensures that $-1 \leq r \leq 1$.
If $r > 0$ the network is said to be assortative, while if $ r < 0$ it is said to be disassortative. If $r = 0$ the network is uncorrelated. Using Equation~(\ref{eqn:Ekq}) and Equation~(\ref{eqn:Phi}), the assortativity in Equation~(\ref{eqn:assort1})  can be written explicitly as
\begin{equation}
    r = \frac{1}{2L} \sum_{i,j}  \frac{A_{ij}(k_i - \mathbb{E}[\kappa]) (k_j -\mathbb{E}[\kappa] ) }{\sigma_\kappa^2}.
    \label{eqn:assortativity}
\end{equation}
This expression highlights that assortativity as the measure of the fluctuations of the degrees of neighbouring nodes around the mean value $\mathbb{E}[k]$. In particular, if $A_{ij} = 1$, the product
\begin{equation}    
    (k_i-\mathbb{E}[\kappa])(k_j-\mathbb{E}[\kappa]),
\end{equation}
shows that for an edge connecting node $i$ to node $j$, if both $k_i>\mathbb{E}[\kappa]$ and $k_j>\mathbb{E}[\kappa]$ or, $k_i<\mathbb{E}[\kappa]$ and $k_j<\mathbb{E}[\kappa]$, the edge will contribute positively to the assortativity. However, edges for which $k_i<\mathbb{E}[\kappa]<k_j$ or $k_j<\mathbb{E}[\kappa]<k_i$ will contribute negatively to the assortativity. Therefore, the assortativity expresses the weighted balance between the number of edges for which both extremities have a degree under or over the mean degree at the end of a randomly chosen edge, and the number of edges for which this mean sits between the degree of the adjacent nodes. 

Additionally, we propose an alternative way to approach the assortativity, beginning with defining the probability $\Psi(K,\Delta)$. This is the probability that the degrees of the nodes at either end of a randomly chosen edge sum to $K$, and differ by $\Delta$. That is, that for a randomly chosen edge, $\kappa + \nu = K$, and $|\kappa - \nu| = \Delta$. The probability $\Psi(K, \Delta)$ can be calculated from the adjacency matrix like so
\begin{eqnarray}
    \Psi(K,\Delta)=\frac{1}{2L} \sum_{ij} A_{ij} \delta_{k_i+k_j,K}\delta_{|k_i-k_j|,\Delta}.
\end{eqnarray} 
The joint average of some measure combining $K, \Delta$, such as their product, is defined as ${\mathbb{E}}[K\Delta]=\sum_{K,\Delta}\Psi(K,\Delta)K \Delta$.
We can show that the assortativity becomes
\begin{eqnarray}
\label{r-K-D}
    r=\frac{\sigma^2_K-\sigma^2_\Delta}{\sigma_K^2+\sigma_\Delta^2},
\end{eqnarray}
where $\sigma^2_K$ and $\sigma^2_\Delta$ are defined as
\begin{eqnarray}
    \sigma^2_K&=&{\mathbb{E}}[K^2]-{\mathbb{E}}[K]^2,\\
    \sigma^2_\Delta&=&{\mathbb{E}}[\Delta^2] - {\mathbb{E}}[\Delta]^2.
\end{eqnarray}
With this picture we now see that the assortativity compares the variance of the marginal distribution $\psi_+(K)$ and $\psi_-(\Delta)$,
\begin{eqnarray}
\psi_+(K)&=& \frac{1}{2L} \sum_{ij} A_{ij}\delta_{k_i+k_j,K}.\label{eqn:K}\\
\psi_-(\Delta)&=& \frac{1}{2L} \sum_{ij}A_{ij}\delta_{|k_i-k_j|,\Delta}.
\label{eqn:psi}
\end{eqnarray}
Plotting this distribution allows one to visualise the assortativity in a different way. Finally, note that the expected statistical error for the assortativity can be calculated using either the jackknife method~\cite{Efron} or the bootstrap method~\cite{Efron_bootstrap}.


\subsection{Generalisation}

\begin{figure}[t!]
    \centering
    \includegraphics[page=7, width=0.45\textwidth]{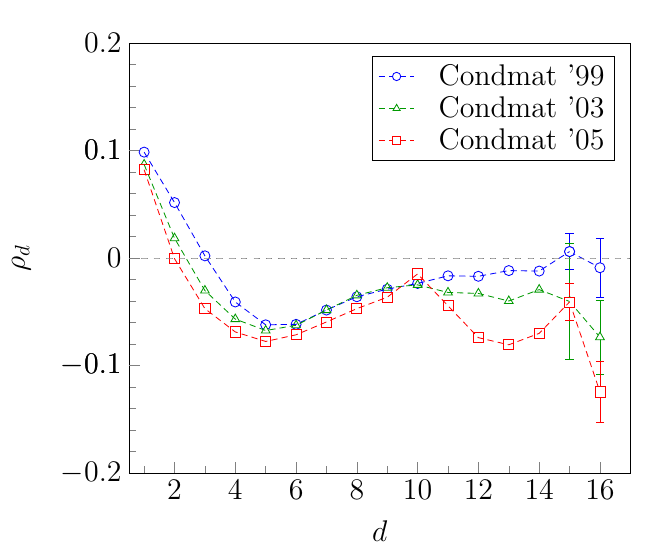} 
    \caption{Correlations in the number of papers per author in the scientific coauthorship network derived from the condensed matter arXiv. Networks are the result of aggregation from the year 1995 to that indicated.}
    \label{fig:correlations_papers}
\end{figure}

In this section, we provide a definition of assortativity beyond nearest neighbours. It is a known result that if we take the adjacency matrix of a network $A$ to a power $n$, then each element of the resulting matrix $A^n_{i, j}$ is the number of paths of length $n$ between nodes $i$ and $j$ \cite{netscience}. We define a modified version of this matrix, $\hat{A}^n$. An element $\hat{A}^n_{i, j}$ of this matrix can either be one or zero. If element $i, j$ of $A^n$ is zero, then so to is the corresponding element of $\hat{A^n}$. The element $\hat{A}^n_{i, j}$ is one if and only if $n$ is the smallest non-negative integer for which $A^n_{i, j}$ is greater than zero. As such, the matrix $\hat{A}^n$ has the following meaning; if $\hat{A}^n_{i, j} = 1$, the nodes $i$ and $j$ are connected by a shortest path of length $n$. If $\hat{A}^n_{i, j} = 0$, no path of length $\le n$ connects the nodes $i$ and $j$. Additionally we use $L_d$ to represent the number of unique pairs of nodes separated by a shortest path of length $d$. Similarly, the degrees of nodes at either end of a randomly chosen path of length $d$ are denoted $\kappa_d$ and $\nu_d$. This allows us to define $\Phi(\kappa_d,\nu_d)$ as
\begin{equation}  
    \Phi(\kappa_d, \nu_d) = \frac{1}{2L_d} \sum_{ij} \hat{A}^d_{i, j} \delta_{k_i \kappa_d} \delta_{k_j \nu_d},
\end{equation}
with $2L_d=\sum_{i,j} \hat{A}^d_{i, j}$. 
The assortativity at a distance $d$ is
\begin{equation}
    r_d = \frac{1}{2L_d} \sum_{ij} \frac{\hat{A}^d_{i, j}\left(k_i - \mathbb{E}[\kappa_d]\right)\left(k_j -\mathbb{E}[\kappa_d]\right)}{\sigma_{\kappa_{d}}^2 },
\end{equation}
with $\sigma_{\kappa_d}^2=\mathbb{E} [\kappa_{d}^2]-\mathbb{E}[\kappa_d]^2$. 
This can be extended to other properties of nodes other than degree as we will demonstrate later (see Figure~\ref{fig:correlations_papers}).
\section{Application}
\begin{figure*}[t!]
    \centering
    \includegraphics[page=1, width=0.45\textwidth]{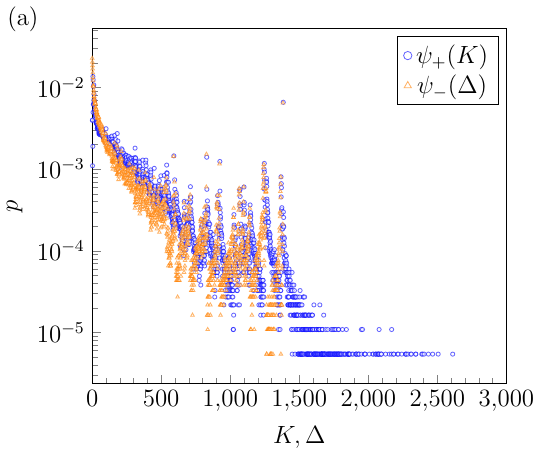} 
    \includegraphics[page=2, width=0.45\textwidth]{Figures/plots.pdf} 
    \includegraphics[page=3, width=0.45\textwidth]{Figures/plots.pdf} 
    \includegraphics[page=4, height=0.378\textwidth, width=0.459\textwidth]{Figures/plots.pdf} 
    \caption{Distributions $\psi_+(K)$ and $\psi_-(\Delta)$, as indicated, of  the sum and difference in degrees of nodes at either extremity of an edge respectively. Panel (a) shows the Enron email network. This network is disassortative and there are numerous edges with a difference of degrees of $\Delta = 1382$. In contrast the PGP web of trust  $\psi_-(\Delta)$ distribution is shown in panel (b). This is assortative and has less noise in the tail of the $\psi_-(\Delta)$ distribution. Panel (c) displays the Facebook users network which is also assortative. Panel (d) however is for the online social network of slashdot users. This network is disassortative and contains a large number of fluctuations as $\Delta$ increases.}
    \label{fig:social_degree_diff}
\end{figure*}

\begin{table*} 
\begin{center}
 \caption{
 The number of nodes is given by $N$, $L$ is the number of edges, $\langle k \rangle $ is the mean degree,  $k_{\text{max}}$ is the largest degree in the network, $\langle k^2 \rangle$ is the second moment of the degree sequence. The average path length is given by  $\ell$ and the diameter by $\ell_{\text{max}}$. The assortativity is denoted by $r$ and the symbol * indicates that the statistical error  is larger than this value indicating there are no degree-degree correlations. }
 \label{tab:social_net}
\small
\centering
  \begin{tabular*}{.98\textwidth}{@{\extracolsep{\fill} }l|r r r r r r r r}

      & $N$ & $L$ & $\langle k \rangle$ & $k_{\text{max}}$ & $\langle k^2 \rangle$ & $\ell$ & $\ell_{\text{max}}$  &  $r$   
      \\ \hline \hline
	  Faux Mesa high school & 147 & 202 & 2.75 & 13 & 11.70 & 6.81 & 16 & 0.12 
	  \\
	  Jazz musicians & 198 & 2,742 & 27.70 & 100 & 1,070.24  & 2.24  & 6& 0.02* 
	  \\ 
	  Infectious diseases & 410 & 2,765 & 13.49 & 50 & 252.43 & 3.63  & 9  & 0.23 
	  \\ 
	  Network science  & 1,461 & 2,742 & 3.75 & 34 & 26.05 & 5.82 & 17 & 0.46
	  \\ 
	  Moreno - health & 2,539 & 10,455 & 8.24& 27& 86.41 & 4.56 & 10 & 0.25 
	  \\
	  Petster & 2,426 & 16,631 & 13.71 & 273 & 582.93 & 3.59 & 10 & 0.05
	  \\
	  PGP web of trust & 10,680 & 24,316 & 4.55 & 205 & 85.98 & 7.49 & 24 & 0.24 
	  \\
	  Cond-mat arXiv & 16,726 & 47,594 & 5.69 & 107 & 73.57 & 6.63 & 18 & 0.19  
	  \\
	  Astrophysics & 18,771 & 198,050 & 21.10 & 504 & 1,379.51 & 4.19 & 14 & 0.21
	  \\	   
	  Twitter & 23,370 & 32,831 & 2.81& 238& 108.17& 6.30 & 15 & -0.48
	  \\	   
	  Google+ & 23,628 & 39,194 & 3.32&2,761 & 1,250.88& 4.03 & 8 & -0.39
	  \\	   	  
	  Munmun & 30,398 & 86,312 & 5.68 & 285 &159.74 & 4.67 & 12 & 0.01
	  \\	   	  
	  Facebook wall & 46,952 & 193,494 &8.24 &223 & 202.87 &5.60 & 18 & 0.25
	  \\	   	  
	  Facebook  & 63,731 & 817,035 & 25.64 & 1,098 & 2,256.80 & 4.32 & 15 & 0.18 
	  \\
	  Slashdot & 79,116 & 467,731 & 11.82 & 2534 & 1,729.86 & 4.04  & 12 & -0.07
	   \\ 
	  Enron email &  87,273 & 299,220 & 6.86 & 1,728 & 1,147.24 & 4.89  & 13 & -0.17 
	   \\\hline	   
  \end{tabular*}
\end{center}
\end{table*}
In this section we analyse the assortativity of 16 different social networks of varying size. We study the marginal distributions given in Equations~(\ref{eqn:psi}) and~(\ref{eqn:K}) for these some of these empirical networks (Facebook, Enron, PGP, and Slashdot. See Figure~\ref{fig:social_degree_diff}), and then calculate the assortativity as a distance for each network.
The datasets we use are (i) the students at Faux Mesa high school friendship network \cite{ResnickFauxMesa,HunterFauxMesa}; (ii) the American jazz musicians from 1912 to 1940 \cite{GleiserJazz}; (iii) the face-to-face contact of participants at an infectious diseases exhibition in Dublin's Science Gallery \cite{IsellaInfectious}; (iv) a coauthorship network of scientists working on network science \cite{netscience}; (v) a Friendship network at Moreno high school \cite{konect:moreno, MoodyMoreno}; (vi) a Friendship network between users of the petster website \cite{konect:petster}; (vii) a key-sharing network from the PGP (Pretty Good Privacy) web of trust \cite{BogunaPGP}; (viii) collaboration networks of authors on the condensed matter \textit{arXiv} from 1995 to 2005 \cite{Newman2001};(ix) the Astrophysics collaboration network \cite{astroph, konect:astroph}; (x) a Twitter user network \cite{twitter_gplus, konect:twitter}; (xi) a Google+ user network \cite{twitter_gplus, konect:gplus}; (xii) a reply network for news site Digg \cite{munmun, konect:munmun}; (xiii) a Facebook wall post network \cite{fbwall, konect:fbwall}; (xiv) a sample of Facebook users in the New Orleans region \cite{ViswanathFacebook}; (xv) a network of users of the technology news site slashdot \cite{KunegisSlashdot,konect:slashdot-zoo}; and (xvi) the network of emails sent at Enron \cite{KlimtEnron,Leskovec}.
All datasets bar (ii), (iii), and (viii) can be found on Konect.cc~\cite{konect:all}. Konect is an open source library of network datasets taken from a wide range of scientific areas.

\begin{table*}
\begin{center}
 \caption{The number of pairs $L_d$ and their corresponding assortativity values $r_d$ up to $d=3$ for each of the 16 networks. The value in parentheses after the assortativity is the error in the last digit calculated using the bootstrap method. For most networks, the assortativity goes to zero (or fluctuates around it) after a small number of steps. A notable exception is that of jazz musicians which becomes more disassortative as $d$ increases.}
\label{tab:correl}
\small
\centering
  \begin{tabular*}{.98\textwidth}{@{\extracolsep{\fill} }l|r r|r r|r r}

    &  $L$ & $r$ & $L_2$ & $r_2$ & $L_3$ & $r_3$  
    \\ \hline \hline
	Faux Mesa high school   &  202  &  0.12 (8) &  410  &  -0.14 (5)  &  561  &  -0.05 (5)
	\\
	Jazz Musicians  &  2,742  &  0.02 (2)  &  10,652  &  -0.13 (1)  &  5,067  &  -0.31 (2)
	\\  
	Infectious diseases &  2,765  &  0.23 (2)  &  13,150  &  -0.00 (1)  &  24,631  &  -0.06 (1)
	\\ 
    Network science & 2,742 & 0.46 (3) & 3,980 & -0.03 (2) & 6,365 & -0.03 (2)
	\\
	Moreno-Health & 10,455 & 0.25 & 71,624 & 0.09 & 365,757 & -0.00 
	\\
	Petster & 16,631 & 0.05 & 222,693 & -0.14 & 757,751 & -0.17
	\\
	PGP web of trust &  24,316  &  0.24 (1)  &  188,183  &  0.01 (1)  &  932,993  &  -0.09 (1)
	\\ 
	Cond-mat arXiv &   47,594  &  0.19 (1)  &  275,120  &  0.07 (1)  &  1,439,255  &  -0.02 (1)
	\\
	Astrophysics & 198,050 & 0.21 & 4,441,041 & -0.08 & 35,788,064 & -0.19 
	\\
	Twitter & 32,831 & -0.48 & 1,110,048 & 0.12 & 2,534,785 & -0.11 
	\\
    Gplus & 39,194 & -0.39 & 12,928,409 & 0.00 & 4,494,516 & -0.01 
    \\
    Munmun-Digg & 86,312 & 0.01 & 2,169,143 & -0.06 & 33,925,631 & -0.14
    \\
	Facebook Wall &  193,494 & 0.25 & 3,149,984 & 0.06 & 32,678,260 & -0.10 
	\\
    Facebook & 817,035 & 0.18 & 31,073,418 & -0.04 & 357,008,740 & -0.14 
	\\
	Slashdot & 467,731 & -0.07 & 47,324,811 & -0.08 & 696,650,899 & -0.11
	\\
	Enron email & 299,220 & -0.17 & 27,781,955 & -0.05 & 282,174,082 & -0.07 \\ \hline
  \end{tabular*}
\vspace{0.5cm}

\end{center}
\end{table*}

\begin{figure*}[t!]
    \centering
    \includegraphics[page=5, width=0.49\textwidth]{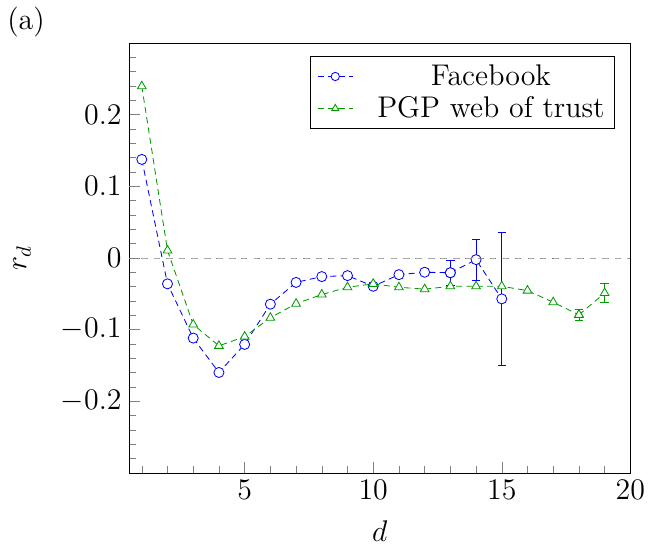} 
    \includegraphics[page=6, width=0.49\textwidth]{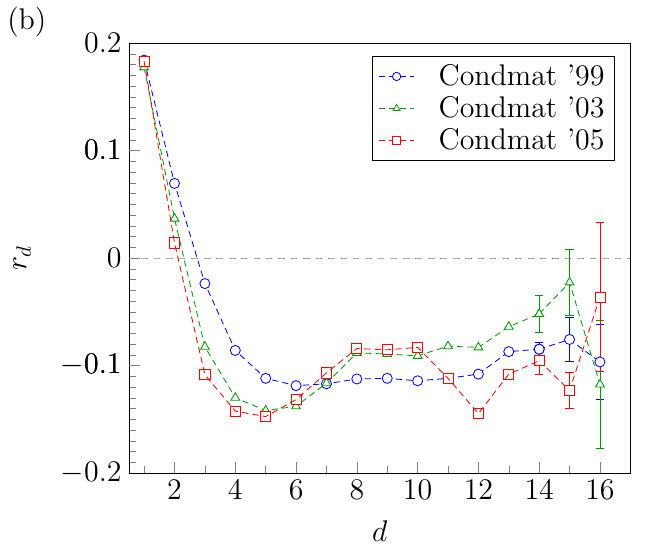} 
    \caption{The degree-degree correlations $r_d$ plotted versus the distance $d$. Panel~(a) shows as $d$ increases for the PGP and Facebook networks, they lose their correlations. Panel~(b) shows the correlations for three time periods on the condensed matter arXiv, which after $d=2$ become anti-correlated. Error bars are calculated using the bootstrap method (they become more prominent at the end due to a significantly smaller number of pairs separated for large $d$).}
    \label{fig:correlations}
\end{figure*}


Some basic properties of these networks are listed in Table~\ref{tab:social_net}. In Refs.~\cite{Newman2002,NewmanPark}, it is observed that social networks tend to be assortative while non-social networks are not. Here, four of these social networks are found to be disassortative, and one is almost uncorrelated. This is in contrast to existing consensus surrounding correlations in social networks. For example, the disassortativity in the network of Marvel Universe characters is noted as being one of the reasons it is dissimilar to real social networks in Ref.~\cite{Gleiser2007}. The Twitter and Google+ networks show the same disassortative behaviour of other large online social networks \cite{HuWang}. The slashdot network also contains `hostile' or `negative' edges. As shown in Refs.~\cite{Szell,MacCarron}, hostile edges contribute towards disassortativity. To examine this further, in the Enron email network we plot the $\psi_{+}(K)$ and $\psi_{-}(\Delta)$ distributions from Equations~(\ref{eqn:K}) and~(\ref{eqn:psi}) in Figure~\ref{fig:social_degree_diff}, and compare them to two assortative networks.

In Figure~\ref{fig:social_degree_diff}~(a), the distributions $\psi_{+}(K)$ and $\psi_{-}(\Delta)$ are shown for the Enron email network. In the tail of the $\psi_{-}(\Delta)$ distribution in particular, we observe a high fraction of edges with $\Delta = 1382$. We can see that the values of $K$ and $\Delta$ are very similar here. Hence the node with the highest degree interacts with multiple nodes which interact with no (or very few) other nodes in the network. This large number of edges between nodes of very different degrees strongly contributes towards the disassortativity.  

The PGP web of trust network on the other hand has a relatively low maximum degree, and low degree variance. It is assortative, and we observe in  Figure~\ref{fig:social_degree_diff}~(b) that there are comparatively few fluctuations in $\Delta$. 

In Figure~\ref{fig:social_degree_diff}~(c-d) the $\psi_+(K)$ and $\psi_-(\Delta)$ distributions are shown for the Facebook and slashdot networks respectively. The Facebook network is assortatively mixed by degree while the slashdot one is not.
There are  large fluctuations around $\Delta \approx 1500$ and $\Delta \approx 2500$ for the slashdot users which drives the disassortativity. However the distribution for Facebook users decays with fewer fluctuations. 

\subsection{Influence}
We turn our attention to correlations at a distance for these 16 social networks.
In Table~\ref{tab:correl} the correlations for pairs of nodes separated up to an edge distance of $d=3$ are displayed. For $d=1$, the number of pairs of nodes $n_{\rm{pairs}}(d)$, and $r_d$, simply correspond to the number of edges and the traditional assortativity of the network, respectively. As $d$ increases, the number of pairs increases rapidly. For most networks, $r_d$ decreases beyond a distance of $d = 1$, and tends to fluctuate about $r_d = 0$ for values of $d$ near the average path length $\ell$. None of the networks have positive correlations at $d=3$, and only two networks have $r_2>0$. 

In Figure~\ref{fig:correlations}~(a), $r_d$ is plotted versus $d$ for two of the larger assortative networks, namely the PGP web of trust and Facebook networks. In each case, the networks are assortative at $d = 1$, with $r_k$ decreasing until $d = 4$, by which point they are disassortative. Beyond $d = 4$, values of $r_k$ plateau below zero. In these networks, users which are a distance of $d = 3$ apart are unlikely to have similar degrees. This is not an unexpected result; as shown in Ref.~\cite{Newman_ego}, an individual's friends are not representative of a social network, instead they are a biased population. Similarly, that reference shows high degree nodes have a disproportionately large effect on networks at a distance of $d=2$. This effect seems to increase with $d$, implying that as distances increase, people become more dissimilar.

The measure we study here is quite different to the degrees of influence observed by Refs.~\cite{C+F2007,C+F2009}. There, the authors study social networks evolving in time, and focus on correlations between node attributes besides degree, such as obesity and happiness. In contrast, we focus on degree correlations, in networks that have either been aggregated in time, or extracted at a snapshot in time. One exception, however, is the condensed matter \textit{arXiv} network, which we examine over multiple time periods; 1995-1999, 1995-2003 and 1995-2005~\cite{Newman2001}. Figure~\ref{fig:correlations}~(b) shows the degree-correlations $r$ versus distance $d$ for these time periods. They each have positive values of $r$ for $d = 1$, which become negative as as $d$ increases.
Although our focus is on degree correlations, one may study correlations as a function of distance for arbitrary node attributes. The scientific coauthorship network, for instance, details the number of papers that node $i$ in the network has published. Studying correlations in this quantity provides a measure of the output assortativity of the collaboration network. These results are shown in Figure~\ref{fig:correlations_papers}, where we plot output correlation, between number of published papers, denoted $\rho_d$, versus the distance $d$. These are correlated initially among nearest neighbours, implying productive authors have productive coauthors, and unproductive authors have unproductive coauthors. Small but nonzero correlations exist at  $d=2$ in the first two time-periods but are decorrelated for the final time point. For $d>2$, there are no positive correlations present. Therefore, on average, two authors three steps away from one another in this coauthorship network are unlikely to have a similar number of publications and this seems to become less similar as time goes on. 
At higher \(d\) values then, authors are more likely to have a different number of publications to their distant neighbours.
\section{Simulations}
Having studied the behaviour of the assortativity for increasing $d$ in empirical networks, we turn our attention to simulated networks. Here we will examine three types of simulated network; Erd\H{o}s-R\'enyi graphs, Erd\H{o}s-R\'enyi graphs  that we have made assortative by rewiring, and configuration model graphs corresponding to each of the empirical networks in Table~\ref{tab:social_net}.
\subsection{Erd\H{o}s-R\'enyi graphs}
Here we generate  Erd\H{o}s-R\'enyi graphs and calculate $r_d$ for all values of $d$ such that $1 \leq d \leq l_{\text{max}}$, where $l_{\text{max}}$ is the diameter of the graph. This is repeated 1,000 times each for Erd\H{o}s-R\'enyi graphs of size $N = 1,000$, with average degree $\langle k \rangle = 5$, 10, and 15. We then repeat this 1,000 times each for graphs of size $N = 10,000$, again with  $\langle k \rangle = 5$, 10, and 15. The results of these calculations are shown below in Figure~\ref{fig:erdosrenyi}.

As we can see, there are no positive correlations between the degrees of nodes at any value of $d$. Negative correlations begin to emerge in all cases at or beyond the average path lengths of the graphs, indicated as dashed vertical lines. This finding is in line with the work of Mayo \textit{et al.}~\cite{Mayo}, who find negative correlations emerge when the average degree is greater than one. While high-degree nodes are exponentially suppressed in Erd\H{o}s-R\'enyi graphs, negative correlations still emerge due to the finite size of the simulations. That is, paths of higher values of $d$ still connect central nodes to nodes at the periphery of the graph. It is reasonable to expect that the degrees of such pairs of nodes would be dissimilar, thus reducing the assortativity at these $d$ values.
\begin{figure*}[t!]
    \centering
    \includegraphics[page=8, width=0.32\textwidth]{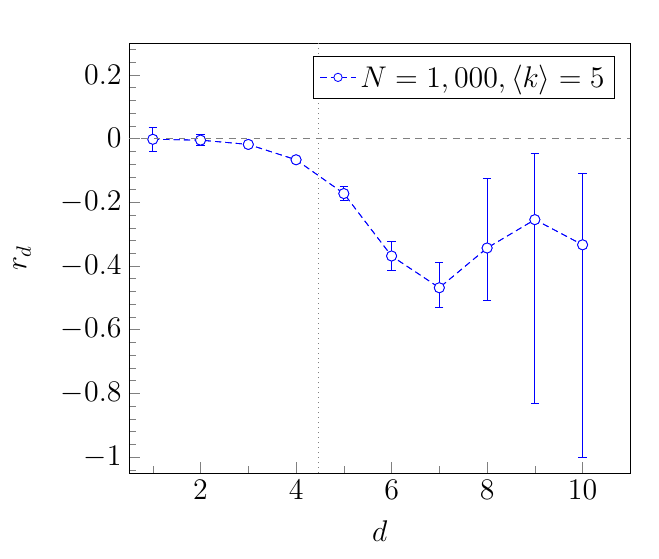} 
    \includegraphics[page=10, width=0.32\textwidth]{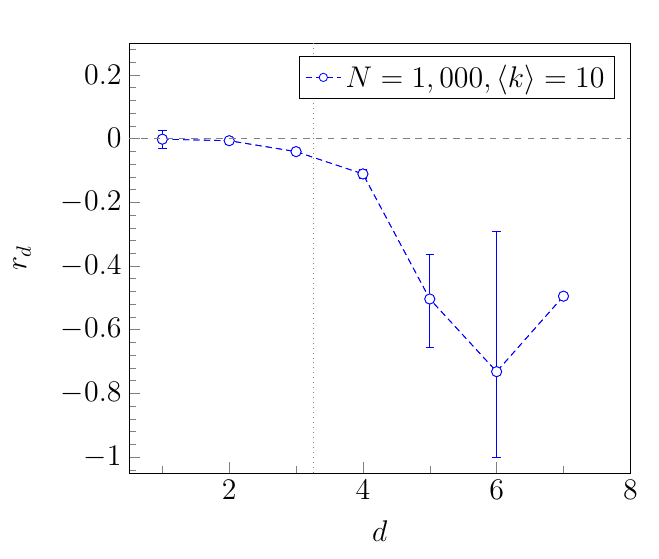} 
    \includegraphics[page=12, width=0.32\textwidth]{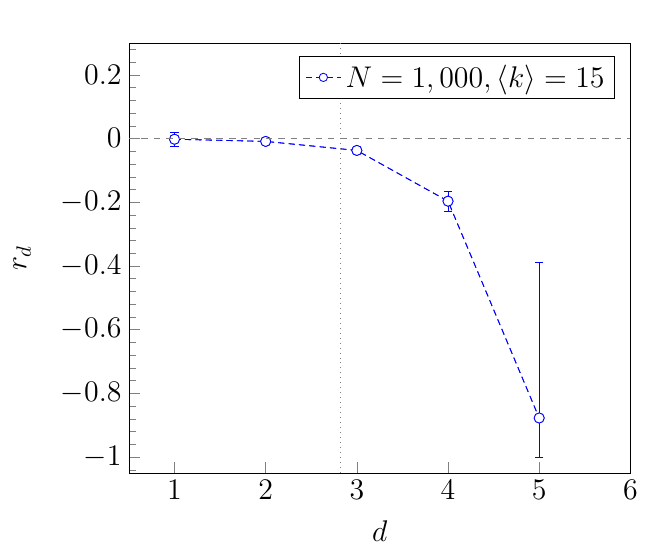} 
    \includegraphics[page=9, width=0.32\textwidth]{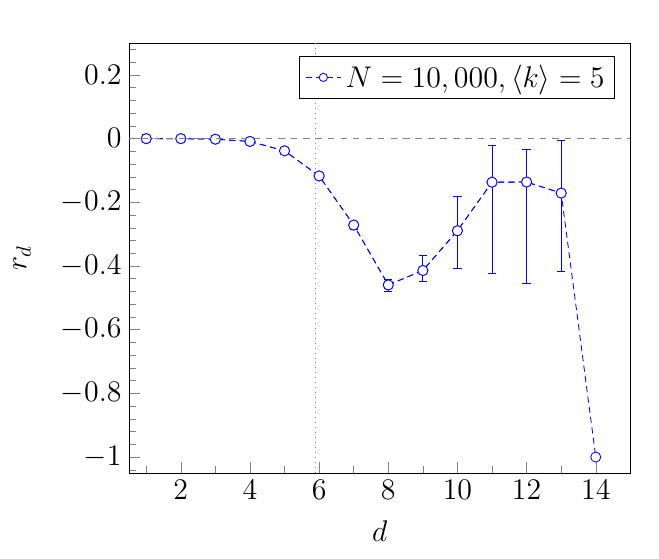} 
    \includegraphics[page=11, width=0.32\textwidth]{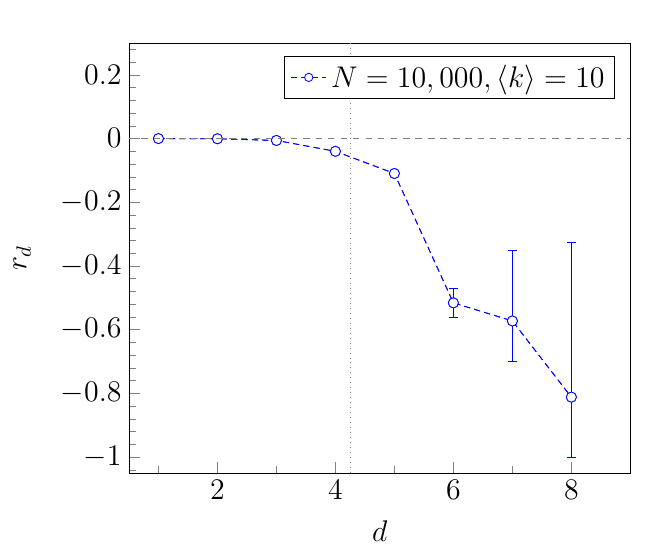} 
    \includegraphics[page=13, width=0.32\textwidth]{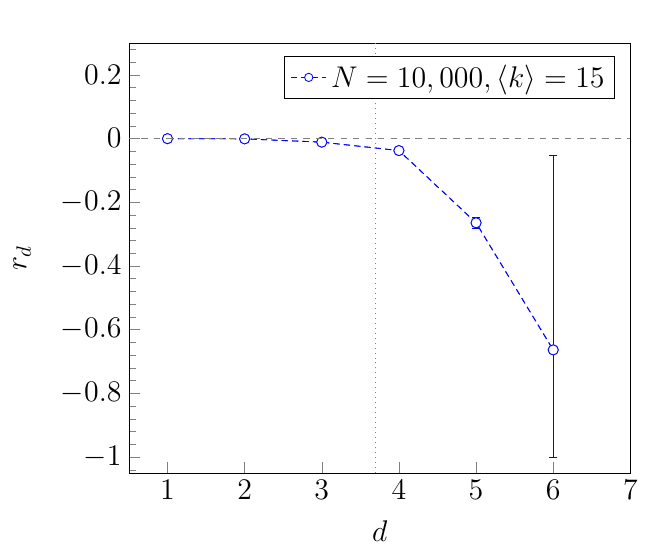} 
    \caption{Results for Erd\H{o}s-R\'enyi graphs with 1,000 nodes (top), and 10,000 nodes (bottom). Points are average values for 1,000 instances of each size and average degree, error bars are 95 percent coverage bars for the assortativity values observed. Vertical dashed lines represent the mean average path length of each set of 1,000 graphs. As we can see in all cases the degrees of connected nodes became negatively correlated as $d$ increases.}
    \label{fig:erdosrenyi}
\end{figure*}

\subsection{Configuration models networks}

\begin{figure}[t!]
    \centering
    \includegraphics[page=14, width=0.9\columnwidth]{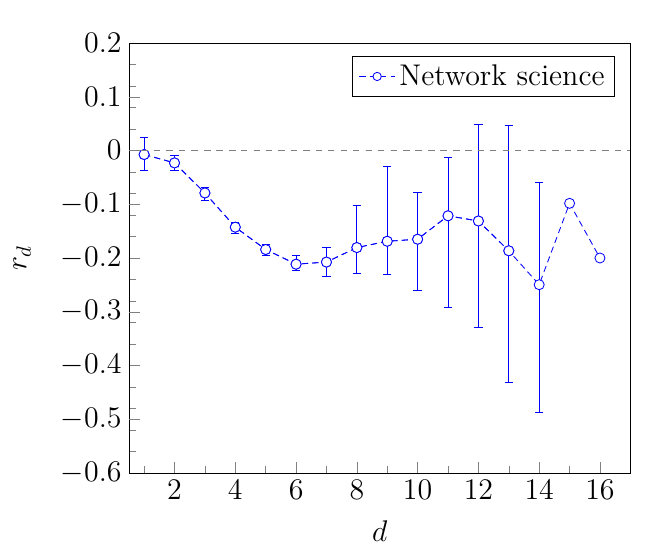} 
    \includegraphics[page=15, width=0.9\columnwidth]{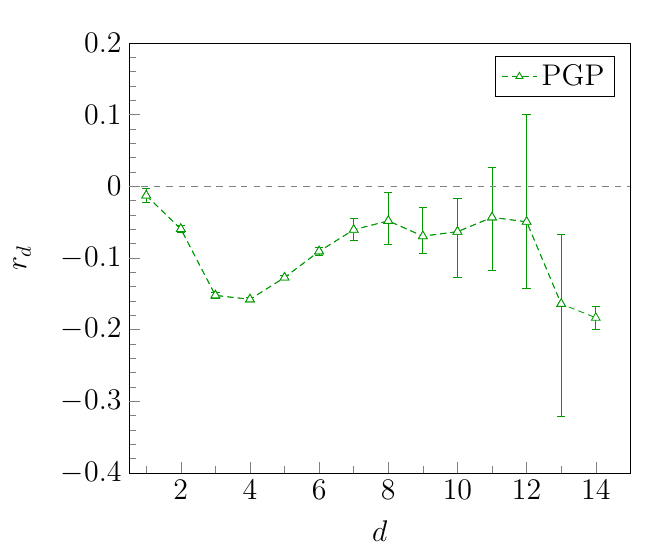} 
    \includegraphics[page=16, width=0.9\columnwidth]{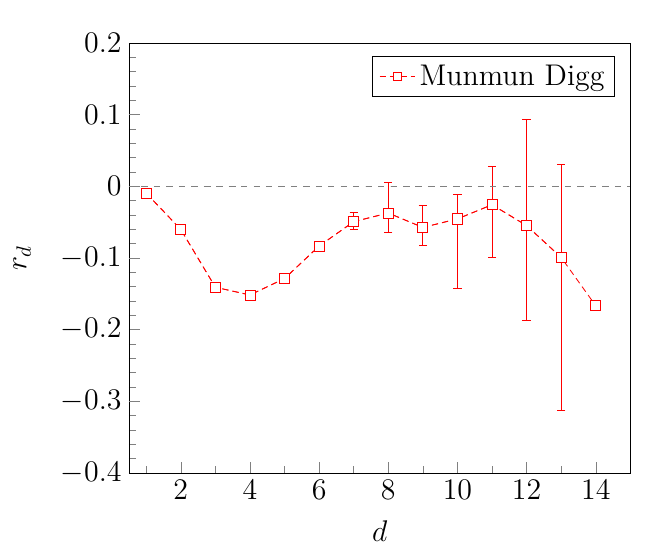} 
    \caption{Degree assortativity $r_d$ versus shortest path length $d$, for configuration model variants of the network science coauthorship network, the PGP web of trust network, and the Munmun-Digg network. Error bars are 95 per cent coverage bars for the assortativity values.}
    \label{fig:correlations_configs}
\end{figure}

We now examine the assortativity as a function of $d$ for configuration model variants of the networks Infectious disease through Munmun-Digg in Table~\ref{tab:social_net}. For each network, 100 instances of the corresponding configuration model network were created, and the assortativity calculated for all $d$ values up to the diameter of each graph. Results for the network science coauthorship, PGP web of trust, and Munmun-Digg networks can be seen in Figure~\ref{fig:correlations_configs}. These results are representative of the remaining networks not shown. 

In all cases we find that the assortativity rapidly decreases before beginning to increase and approach zero for higher values of $d$. In the three cases shown here we also see strong negative correlations at the largest values of $d$ for each network. These spikes are due to the fact that at these large $d$ values, there are very few pairs separated by shortest paths of that particular length $d$.

One possible explanation for the decrease in assortativity is that the quantity reaches its minimum when the influence of hubs in the network is at its maximum. For example, high-degree nodes will be connected to many other nodes by shortest paths of lengths 2 and 3, and these nodes will have degrees that are dissimilar to that of the hub node, thus reducing the assortativity. As $d$ increases further,the number of pairs of nodes separated by shortest paths of length $d$ increases. Many of these pairs then will be positively assortative, which counteracts the effect of the hub nodes.

\begin{figure}[t!]
    \centering
    \includegraphics[page=29, width=0.9\columnwidth]{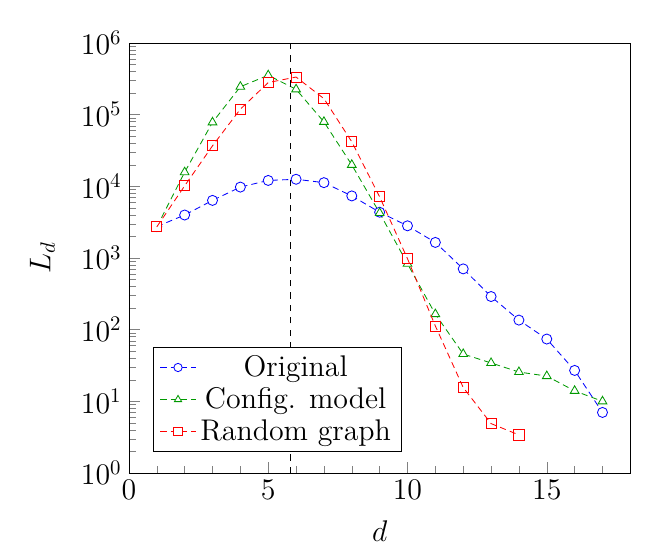} 
    \includegraphics[page=30, width=0.9\columnwidth]{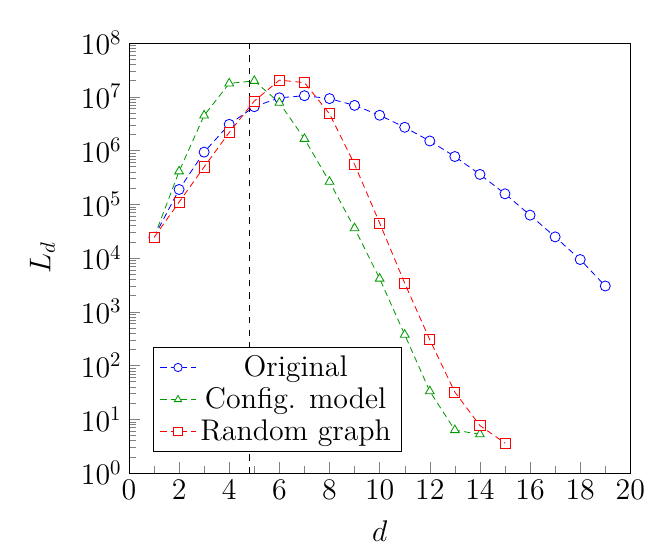} 
    \caption{The number of pairs for each shortest path length $d$ for the Network Science coauthors (top) and PGP web of trust (bottom). The dashed vertical line represents the average path length of the original network.}
    \label{fig:npairs}
\end{figure}

To illustrate this further, we plot the number of pairs at each distance $d$ in Figure~\ref{fig:npairs}. Here we see after the average path length (dashed vertical line), these is a sharp drop in the number of pairs. Beyond this point, the hubs have probably reached every periphery node, but nodes at different ends of the network still not be connected. As the hub is likely to be at a different side of $\mathbb{E}[\kappa]$ to a periphery node in Equation~(\ref{eqn:assortativity}), their influence will contribute negatively to the assortativity. This difference in degree for neighbours can be seen in Figure~\ref{fig:social_degree_diff}, these spikes will only increase as the number of paths increase.

\subsection{Modified Erd\H{o}s-R\'enyi graphs}
\begin{figure*}
    \centering
    \includegraphics[page=17, width=0.32\textwidth]{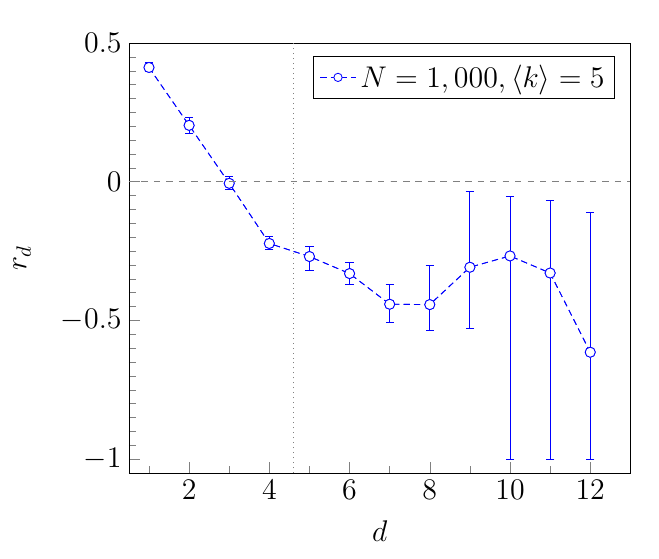} 
    \includegraphics[page=19, width=0.32\textwidth]{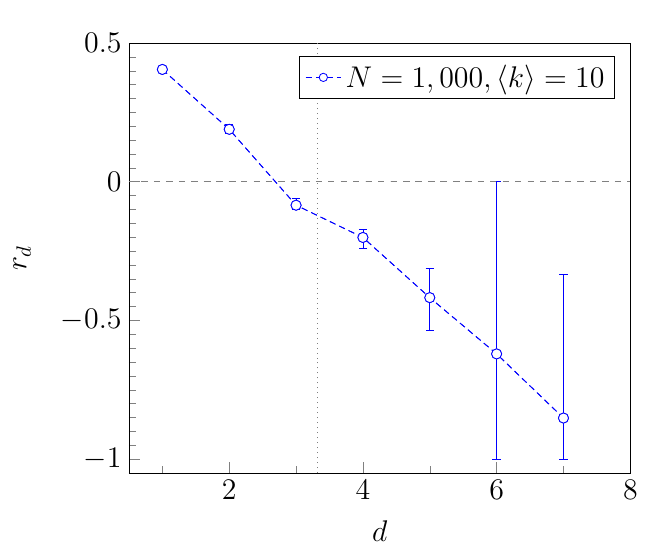} 
    \includegraphics[page=21, width=0.32\textwidth]{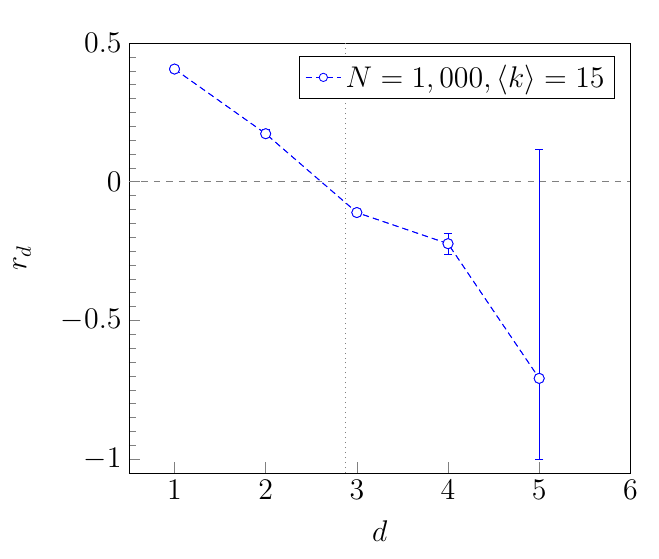} 
    \includegraphics[page=18, width=0.32\textwidth]{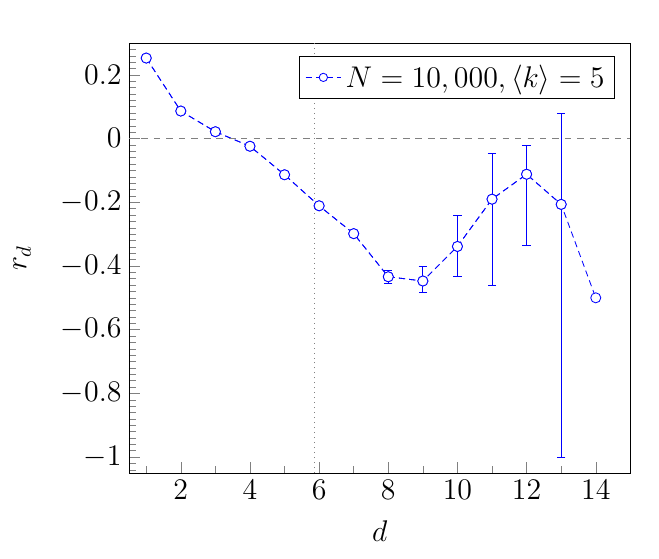} 
    \includegraphics[page=20, width=0.32\textwidth]{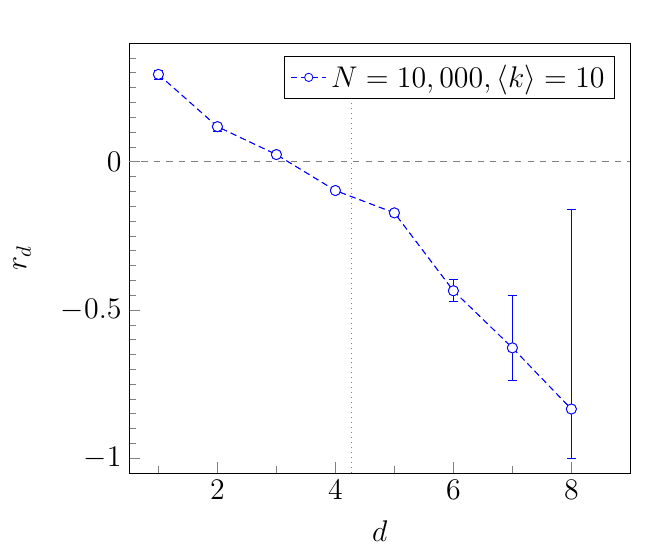} 
    \includegraphics[page=22, width=0.32\textwidth]{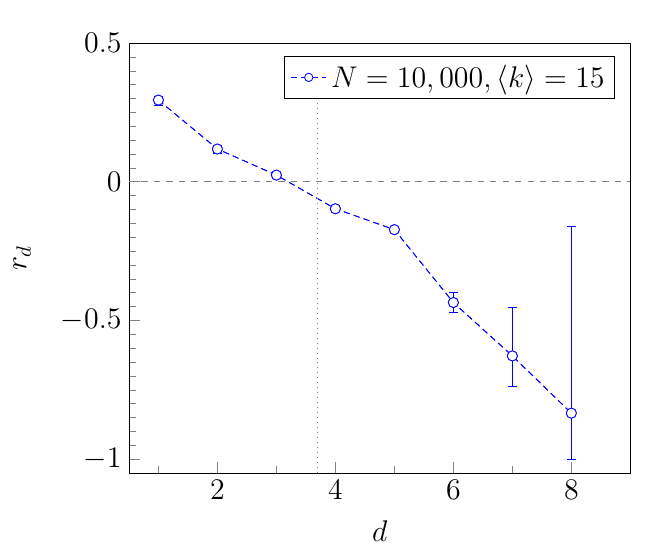} 
    \caption{Correlations between the degrees of nodes separated by shortest paths of length $d$ for Erd\H{o}s-R\'enyi graphs of size 1,000 nodes (top, 1,000 instances for each value of $\langle k \rangle$), and 10,000 nodes (bottom, 500 instances each for $\langle k \rangle = 5$ and $\langle k \rangle = 10$, 250 instances for $\langle k \rangle = 15$) that have been rewired to be assortative. Vertical dashed lines represent the mean average path length of each set of 1,000 graphs. The error bars are 95 per cent coverage bars for the assortativity values.}
    \label{fig:correlations_configs2}
\end{figure*}

Here we start with Erd\H{o}s-R\'enyi graphs of size $N = 1,000$ and average degree $\langle k \rangle = 5$, 10, and 15. We run a rewiring algorithm on 1,000 instances of each graph and calculate the assortativity for all values of $d$ up to the diameter of the rewired graph. The algorithm works by randomly removing edges that decrease the assortativity, and replacing them with edges that increase the assortativity, as per Equation~\ref{eqn:assortativity}. This process is repeated until the desired value of first-neighbour assortativity is reached, choosing $r = 0.4$ for graphs size $N = 1,000$ and $r = 0.25$ for graphs of size $N = 10,000$. The results for this process are shown in Figure~\ref{fig:correlations_configs2}. This approach is then repeated $500$ times for Erd\H{o}s-R\'enyi graphs of size $N = 10,000$ and average degrees $\langle k \rangle = 5$ and 10, and 250 times for Erd\H{o}s-R\'enyi graphs of size $N = 10,000$ and average degree $\langle k \rangle = 15$. The reduced number of iterations here is due to the increased computational cost of rewiring larger graphs. As the number of nodes and edges increases, the contribution that each edge makes to the overall assortativity decreases and so many more iterations of the rewiring algorithm are required to reach the desired assortativity value.

As we can see, all of the graphs started out assortative at $d = 1$ due to the rewiring algorithm, but these positive correlations quickly disappear as $d$ increases, and beyond the average path lengths, the results are very similar to those seen for the Erd\H{o}s-R\'enyi graphs in fig~\ref{fig:erdosrenyi}. This is not a great surprise, but serves to highlight that the degree distribution plays a role the behaviour of the decay of the assortativity as the value of \(d\) increases. 
\section{Conclusions}
Social networks are generally found to be assortatively mixed by degree, a property separating them from other complex networks. 
Here we extend this to go beyond nearest neighbours in order to test if they remain assortative or if this changes as the distance increases.

We find that for all cases, as the edge distance between nodes increases, degree-degree correlations vanish. In some cases, they become negatively correlated before becoming uncorrelated. In all 16 social networks, nodes three steps away from each other are more likely to have dissimilar degrees. That is, nodes three steps away from each other are more likely to have degrees on either side of $\mathbb{E}[\kappa]$. 
The same is true of number of papers in the condensed matter arXiv coauthorship network, and in that dataset, over time, both quantities become less correlated.

We performed simulations on three different types of network to test this. With Erd\H{o}s-R\'enyi random graphs, correlations remain close to zero as expected from mean-field arguments for distances $d$ up to the average path length. Beyond this, while correlations become negative, there are fewer pairs corresponding to these distances, and fluctuations are large as a result. Further, this seems to be a reflection of the finite size of the simulations, with pairs of distant nodes tending to consist of one high-degree node, and one peripheral node at the end of a chain. In contrast, configuration model variants of real-world networks, correlations quickly become negative as $d$ increases. Finally in Erd{\H o}s-R{\'e}nyi graphs rewired for assortativity, correlations become negative for distance beyond the average path length, as before. This is due to their similarity to Erd\H{o}s-R\'enyi random graphs, despite rewiring.

We attribute the decreasing degree correlations to the right-skewed distributions of the empirical datasets and, hence, the influence of hubs in these networks. Hubs are nodes with large degrees, and due to the initial assortativity in social networks, they are in general more likely to connect to nodes with a similar degree. However, they will have quite different degrees to others as the distance increases. 

With the calculation for the assortativity, if the degrees are on
either side of the average degree at the end of an edge
this contributes negatively towards the assortativity value. Therefore, as the distance increases, a pair is more likely to contain a well connected node and one at the periphery. 
As the number of pairs increases significantly at each step, these high degree nodes interact more and more with low degree nodes. However, by a distance around the average path length, a hub may be connected to every node in the graph. Beyond this distance then, the influence of hubs on the assortativity value is diminished.

To investigate this further, we next aim to explore the role of the degree distribution with beyond nearest-neighbour correlations and identify how much influence hubs play on this. We also aim to explore more of the behaviour of disassortative networks, both in empirical data and simulations. Finally the assortativity at a distance here is calculated from the point of view of one shortest path between a pair, this could be weighted by the number of shortest paths, this is another extension we intend to explore.

In Ref.~\cite{Mayo}, using an average degree approach for three social networks it was also shown that the assortativity decreases as the distance increases. Here we show that for more social networks, as well as simulated networks, this is always the case. Similarly on a network at different time points, both degree assortativity and correlations between the number of papers of connected authors become negative at distances above two. These negative correlations get stronger as time increases. This result is the opposite of what the ``three degrees of influence'' work would suggest. Our work instead implies, that in a social network, the more distant you are from someone, the more likely you are different to them. This is related to the idea of the ``friendship paradox'' -- that your friends, on average, have more friends than you -- and is also likely because social networks are right-skewed.

This again implies that our observations of the correlations between nodes separated by distances greater than one here are tied to the degree distributions of the graphs, and the future work outlined above will help to explain this connection.
\section*{Funding}
SM is funded by the Science Foundation Ireland under Grant number 18/CRT/6049.

\section*{Acknowledgment}
\vspace{-0.5cm}
The authors would like to thank Ralph Kenna for ideas on assortativity and conceptual discussions.

\end{document}